\documentclass[conference,onecolumn]{IEEEtran}
\usepackage[utf8]{inputenc}
\usepackage[bookmarks]{hyperref}
\hypersetup{colorlinks=true,citecolor=blue,linkcolor=blue,filecolor=blue,urlcolor=blue}

\usepackage{fullpage}
\usepackage{tikz}
\usepackage{circuitikz}

\usepackage{amsmath,amsfonts,amssymb,amsthm,mathtools}

\usepackage{cleveref}

\crefname{thm}{Theorem}{Theorems}
\crefname{prop}{Proposition}{Propositions}
\crefname{lem}{Lemma}{Lemmas}
\crefname{cor}{Corollary}{Corollaries}
\theoremstyle{definition}
\crefname{ref}{Remark}{Remarks}

\usepackage[hang,flushmargin]{footmisc}
\usepackage{tikz}
\usepackage{pgfplots}

\usepackage{xifthen} 

\newcommand{\cX}{\mathcal{X}}
\newcommand{\cY}{\mathcal{Y}}

\newcommand{\cP}{\mathcal{P}}

\newcommand{\CeaO}[1][]{
	\ifthenelse{\isempty{#1}}
	{\cC^{(1)}_{\mathrm{ea}} }
	{\cC^{(1)}_{\mathrm{ea},#1}}
}

\begin{document}
\title{A Tight Uniform Continuity Bound for Equivocation}

\author{%
  \IEEEauthorblockN{{Mohammad A.~Alhejji}
  \IEEEauthorblockA{JILA, University of Colorado/NIST \\
                    440 UCB, \\
                    Boulder, CO 80309, USA\\
                    Email: mohammad.alhejji@colorado.edu}}
  \and
  \IEEEauthorblockN{Graeme~Smith}
 \IEEEauthorblockA{JILA, University of Colorado/NIST\\
                    440 UCB,\\
                    Boulder, CO 80309, USA\\
                    Email: graeme.smith@colorado.edu}
}

% use for special paper notices
%\IEEEspecialpapernotice{(Invited Paper)}

% make the title area
\maketitle

\begin{abstract}
\begin{center}
 We prove a tight uniform continuity bound for the conditional Shannon entropy of discrete finitely supported random variables in terms of total variation distance. 
\end{center}
\end{abstract}

\IEEEpeerreviewmaketitle

\section{Introduction}

\IEEEPARstart{T}{here} has been significant work towards understanding the interplay between the Shannon entropy and various distance measures on the probability simplex \cite{Z07, HY10, S13}. In conjunction, remarkable progress has been made in the study of the continuity of the von Neumann entropy, or matrix entropy, versus trace distance \cite{F73,A07,HN17}. In both cases, several proofs of a tight uniform continuity bound have been presented \cite{Z07, HY10, P18, A07,W16,HN17_b}. On the other hand, uniform bounds, which are not tight but are independent of the size of the conditioning system, were proven for the conditional Shannon and von Neumann entropies in \cite{AF04, W16}. In this note, We present a proof of a tight uniform continuity bound for the conditional Shannon entropy of discrete finitely supported random variables in terms of total variation distance.

Let $\cX := \{1,2,$$..., |\cX|\}$ and $\cY := \{1,2,$$..., |\cY|\}$. Let $\cP_{\cX \times \cY}$ denote the probability simplex with $|\cX| |\cY|$ atoms. The Shannon entropy of a random variable $X \sim p_{X}$ is defined as \cite{C91}
\begin{align}
    H(X) := - \sum_{i \in \cX} p_{X}(i) \log{p_{X}(i)}.
\end{align}
Here, all logarithms are meant in base 2. Similarly, for two jointly distributed random variables $(X,Y) \sim p_{XY}$, the joint Shannon entropy is given by
\begin{align}
    H(XY) := - \sum_{i \in \cX} \sum_{j \in \cY} p_{XY}(i,j) \log{p_{XY}(i,j)}, 
\end{align}
and the conditional Shannon entropy $H(X|Y)$, also known as the equivocation in $X$ given $Y$, is defined as
\begin{align}
    H(X|Y) :&= H(XY) - H(Y) \label{eq: equiv1} \\
    &=\sum_{j \in \cY} p_{Y}(j) H(X|Y=j) \label{eq: equiv2} \\
    &= - \sum_{i \in \cX} \sum_{j \in \cY}  p_{XY}(i,j) \label{eq: equiv3} \log{\frac{p_{XY}(i,j)}{p_{Y}(j)}}.
\end{align}
Given two probability distributions $p_{XY},q_{X'Y'} \in \cP_{\cX \times \cY}$, the total variation distance between them is given by
\begin{align}
    \text{TV} (p_{XY}, q_{X'Y'}) = \frac{1}{2} \sum_{i \in \cX} \sum_{j \in \cY} | p_{XY}(i,j) - q_{X'Y'}(i,j)|.
\end{align}
A bona fide metric on the probability simplex, it captures the distinguishability of any two probability distributions \cite{FG99}.

There are at least two reasons to want the sharpest bounds on the variation of entropic quantities in terms of relevant distance measures. The first reason is that in practice we only have access to estimates of distributions, never the true distribution, of any given random variable. Second, in settings where practical algorithms for computing communication rates are known only for a special class of distributions, it pays to have tight bounds on the error incurred by approximating the rates of arbitrary distributions by the ones of the closest members of said special class  \cite{LG09, SH17, SSWR17}. \\

\noindent \textit{Main Result:}
Let $\epsilon \in (0,1 - \frac{1}{|\cX|}]$ and $p_{XY},q_{X'Y'} \in \cP_{\cX \times \cY}$ be such that $\text{TV} (p_{XY}, q_{X'Y'}) \leq \epsilon$. The following inequality holds
\begin{align}\label{eq:bound}
&|H(X|Y) - H(X'|Y')| \leq
    \epsilon \log{(|\cX| -1)} + h(\epsilon),
\end{align}
where $h(\cdot)$ is the binary entropy function. Moreover, the inequality is tight. This means that for all $\epsilon \in (0,1 -\frac{1}{|\cX|}]$, there exists a pair of probability distributions on $\cX \times \cY$ such that the total variation distance between them is $\epsilon$ and \eqref{eq:bound} is saturated.

The mechanics of the proof are based on $G$-majorization, which refers to a group-induced preordering on a vector space \cite{MOA79,S90}. It is a generalization of Schur majorization, where the group in question is the symmetric group. That is, we say that the vector $q_{X'Y'}$ majorizes the vector $p_{XY}$ whenever $p_{XY}$ is in the convex hull of the orbit of $q_{X'Y'}$ under the action of the symmetric group, where the action is thought of in terms of the natural permutation representation. Since the Shannon entropy is strictly concave and invariant under the action of the symmetric group, this implies $H(XY) \geq H(X'Y')$, i.e., the Shannon entropy is strictly Schur-concave. 

Equivocation is not Schur-concave. While it is concave, it is not invariant under the action of the symmetric group. It is, however, invariant under the action of a proper subgroup of the symmetric group. We use this invariance to prove the present result.

\section{Proof}

Consider the set $\{p_{XY}(i,j) \: |\:  i \in \cX; j \in \cY\}$, where $p_{XY}$ is a bivariate probability distribution. The conditional entropy $H(X|Y)$ associated with $p_{XY}$ is invariant under permutations of $j$ indices, as can be seen from \eqref{eq: equiv2}. Additionally, it is invariant under exchanges of the form $(i_{1},j) \leftrightarrows (i_{2},j) \: \forall \:  i_{1},i_{2} \in \cX$ and $j \in \cY$. These permutations generate the symmetries of the equivocation in $X$ given $Y.$ Denote this subgroup of the symmetric group by $S_{\cX|\cY}$. 

% We note that the bound is monotonic in $\epsilon$ for $\epsilon \in (0,1 - \frac{1}{|\cX|}]$, and so we are free to take $\text{TV} (p_{XY}, q_{X'Y'}) = \epsilon$. 
We assume without loss of generality that $H(X|Y) \geq H(X'|Y')$. We proceed in three steps.

\subsection{Reordering}
First, we order the components of the bivariate probability distributions in a suggestive way. We arrange the components to be in blocks of $|\cX|$ components that share the same $Y$ label. For definiteness, we order the $|\cY|$ blocks such that $q_{Y'} (j)  - p_{Y} (j)$ is non-increasing in $j$. For the block labeled by $j$, we define the two sets:
\begin{align}
    I_{j} = \{ i \: | \: q_{X'Y'} (i,j) \geq p_{XY} (i,j) \}, \\
    I_{j}^{c} = \{ i \: | \:   q_{X'Y'} (i,j) < p_{XY} (i,j) \}.
\end{align}
Remark that either of the two sets could be empty, but not both. Within the $j$th block, if both $I_{j}$ and $I_{j}^{c}$ are nonempty, put the elements of $I_{j}$ ahead of those in $I_{j}^{c}$. That is, we permute the components in both vectors simultaneously so that $i < i' $ for all $ i \in I_{j}$ and $ i' \in I_{j}^{c}$. Next, we permute the components so that $q_{X'Y'} (i,j)$ is non-increasing in $i$ for $i \in I_{j}$. We do the same for the components associated with the elements of $I_{j}^{c}$. Note that in addition to preserving total variation distance, all of these operations are in $S_{\cX|\cY}$ and so, they do not affect the equivocations.

\subsection{Walking}
The second step involves an optimized form of a proof technique due to Pinelis (see the third answer here \cite{P18}). We walk the two probability distributions across the probability simplex until $H(X'|Y') = 0$. In the process, we take care that the difference of equivocations does not decrease and that the total variation distance does not increase.

We start by zooming into the $j$th block. If $I_{j}$ is not empty, then we make the following replacements: 
\begin{align}
     &q_{X'Y'} (1,j) \mapsto q_{X'Y'} (1,j) + [q_{X'Y'} (i,j) - 
      p_{XY} (i,j)], \\
     &q_{X'Y'}(i,j) \mapsto  p_{XY}(i,j),
\end{align}
consecutively for each $i \in I_{j} \setminus{\{1\}}$. If $I_{j}$ consists of a single element, then such replacements need not be done. By design, these replacements do not affect total variation distance. To see that $H(X'|Y')$ did not increase, we note that the old probability vector is in the convex hull of the orbit of the new vector under the action of $S_{\cX | \cY}$. Put another way, the probability weights, conditional on $j$, are now no less concentrated than before. After these replacements are made, the following inequalities will hold:
\begin{align} 
    &q_{X'Y'} (1,j) - p_{XY} (1,j) \geq 0, \label{eq: form}\\
    &p_{XY} (i,j) - q_{X'Y'} (i,j) \: \geq 0, \label{eq: form1}
\end{align}
for each $i \in \cX \setminus{\{1\}}$. 

Next, we make $q_{X'Y'} (1, j) = q_{Y'} (j)$ by transferring probability weights in the block from the bottom to the top. Specifically, we make the following replacements:
\begin{align}
     &q_{X'Y'} (1,j) \mapsto q_{X'Y'} (1,j) + q_{X'Y'}(i,j), \\
     \quad &p_{XY} (1,j) \mapsto p_{XY} (1,j) + q_{X'Y'}(i,j), \\
     &q_{X'Y'}(i,j) \mapsto  q_{X'Y'}(i,j) - q_{X'Y'}(i,j), \\
     \quad &p_{XY}(i,j) \mapsto  p_{XY}(i,j) - q_{X'Y'}(i,j),
\end{align}
consecutively for each $i \in \cX \setminus{\{1\}}$. Note that the transfers are made in both probability distributions to ensure that the total variation distance remains the same. Observe what happens when a probability weight $s$, suitably small and non-negative, is taken from outcome ${(i,j)}$ and given to outcome ${(1,j)}$ in both probability distributions. The difference of entropies changes by the following amount:
\begin{align*}
    &\{ [ \eta ( q_{X'Y'} (1,j) + s) - \eta ( p_{XY} (1,j) + s)] \\
    & - [ \eta ( q_{X'Y'} (1,j)) - \eta ( p_{XY} (1,j))]\} \\
     &+ \{ [ \eta ( p_{XY} (i,j)) - \eta ( q_{X'Y'} (i,j))] \\
     & - [ \eta ( p_{XY} (i,j) - s) - \eta ( q_{X'Y'} (i,j) - s)]\},
\end{align*}
where $\eta(x)=x\log{x}$. Since it is convex and inequalities \eqref{eq: form} and \eqref{eq: form1} hold, the two differences in curly brackets are not negative. Hence, we can set $q_{X'Y'} (1,j) = q_{Y'} (j)$ and $q_{X'Y'} (i,j) = 0$ for $i \neq 1$. This implies that $q_{Y'} (j) H(X'|Y' = j)$ vanishes.

Now, say $I_{j}$ is empty. Then it holds that $q_{X'Y'} (i,j) - p_{XY} (i,j) < 0$, for all $i \in \cX$. It also holds that $q_{X'Y'} (1,j) \geq q_{X'Y'} (2,j) \geq ... \geq q_{X'Y'} (|\cX|,j)$. Our goal is still to take $q_{Y'} (j) H(X'|Y' = j)$ to zero. To do this, we increase $q_{X'Y'} (1,j)$ by transferring in probability weights from the rest of the outcomes in the block. As explained before, such transfers can only decrease $H(X'|Y')$. Furthermore, as long as $q_{X'Y'} (1,j) - p_{XY} (1,j) < 0$, these transfers can be done without affecting the total variation distance. If at any point $q_{X'Y'} (1,j) - p_{XY} (1,j) = 0$, we stop as \eqref{eq: form} and \eqref{eq: form1} now hold for this block. In such a case, we start the process mentioned in the previous paragraph. Otherwise, we keep going until $q_{X'Y'} (1,j) = q_{Y'} (j)$.

After all blocks have been processed, we can assume without any loss of generality that $q_{X'Y'} (i,j) = 0$ for all $i \neq 1$. With this in mind, we subject both probability distributions to a stochastic process that averages over all the blocks. Specifically, 
\begin{align}
    \mathcal{E} : \nu_{XY} (i,j) \mapsto \frac{1}{|\cY|} \sum_{j \in \cY} \nu_{XY} (i,j). 
\end{align}
The stochastic map $\mathcal{E}$ is a convex combination of elements in $S_{\cX | \cY}$. This implies that $H(X|Y)$ will not decrease. Of course this processing does not change the fact that $H(X'|Y') = 0$ since only $q_{X'Y'}(1,j)$ can be nonzero. Recall that the total variation distance between two distributions does not increase under stochastic maps. Note that the outputs of $\mathcal{E}$ are always product distributions with a uniform marginal on $\cY$. As for $\cX$, we have the following marginals
\begin{align}
  q_{X'} (1) &= 1  \: \geq \: p_{X} (1) \: \geq \: 1 - \epsilon\label{final1}\\
  q_{X'} (i) &=  0  \: \leq \: p_{X} (i), \label{final}
\end{align}
for $i \in \cX \setminus{\{1\}}$.

\subsection{Estimating}
Since the two distributions can be assumed to be product distributions on $\cX \times \cY$, then from here one can invoke the bound for the unconditional case \cite{Z07} to finish the proof. For the sake of completeness, we include the following rather standard estimates. 

Given distributions as in \eqref{final1} and \eqref{final}, we can upper-bound $H(X)$ in the following way:
\begin{align*}
    H(X) &=  -p_{X}(1) \log{p_{X}(1)} 
     - \sum_{i \neq 1} {p_{X}(i)} \log{{{p_{X}(i)}}} \\
    & \leq  -p_{X}(1) \log{p_{X}(1)}  \\
    &\quad - \sum_{i \neq 1} {\frac{(1 - p_{X}(1))}{|\cX| - 1}} \log{{{\frac{(1 - p_{X}(1))}{|\cX| - 1}}}}\\
    & = (1 - p_{X}(1)) \log{(|\cX| - 1)} + h((1 - p_{X}(1)))\\
    & \leq \epsilon \log{(|\cX| - 1)} + h(\epsilon)
\end{align*}
where the first inequality follows because the vector $(\frac{1 - p_{X}(1)}{|\cX| - 1})_{i=2}^{|\cX|}$ is (Schur) majorized by $(p_{X}(i))_{i=2}^{|\cX|}$. The second inequality comes from the monotonicity of the bound. This completes the proof. 

To see that the bound is tight, let $\epsilon \in (0, 1 - \frac{1}{|\cX|}]$ be given and consider the following probability distributions.
\begin{align}
    q_{X'Y'}(1,1) &= 1 \\ 
    p_{XY}(1,1) &= 1 - \epsilon \quad \text{and} \quad p_{XY}(i,1)= \frac{\epsilon}{|\cX|-1}, 
\end{align}
for $i \in \cX \setminus{\{1\}}$. Evidently, the two are separated by $\epsilon$ in total variation distance and the associated equivocations saturate the bound.

\section{Concluding remarks}
We have presented a proof of a tight uniform continuity bound for the conditional Shannon entropy. The bound is independent of the alphabet size of the conditioning system. However, we have assumed in the proof that the conditioning system has finite support. It would be interesting to prove the bound without such an assumption, but we leave that as an open problem. 

The proof depends crucially on the invariance of $H(X|Y)$ under the action of $S_{\cX | \cY}$. Various forms of conditional Rényi entropies share this invariance \cite{T15, A77}. More generally, entropic quantities of interest are invariant under the action of some subgroup of the symmetric group. For example, the mutual information $I(X;Y) = H(X) + H(Y) - H(XY)$ is invariant under
"local" permutations of indices, i.e., tensor products of permutations of $X$ and $Y$ labels. We believe such symmetries provide a path towards a better understanding of the behaviors of the corresponding quantities in terms of continuity and beyond.

\section*{Acknowledgment}
We thank Shawn Geller, Alexander Kwiatkowski, Felix Jimenez, Felix Leditzky and Emanuel Knill for useful discussions and insightful comments regarding the manuscript. GS was supported by NSF CAREER award CCF 1652560.

\bibliographystyle{IEEEtran}
\bibliography{IEEEabrv,references.bib}

\end{document}